\documentstyle[12pt]{article}  
\newlength{\extralineskip}
\newcommand{\beq}{\begin{equation}}
\newcommand{\eeq}{\end{equation}}
\newcommand{\bd}{\begin{displaymath}}
\newcommand{\ed}{\end{displaymath}}

\def\bea{\begin{eqnarray}}
\def\eea{\end{eqnarray}}

\def\ba{\beq\new\begin{array}{c}}
\def\ea{\end{array}\eeq}

\def\inbar{\,\vrule height1.5ex width.4pt depth0pt}
\def\IC{\relax\hbox{$\inbar\kern-.3em{\rm C}$}}
\def\IR{\relax{\rm I\kern-.18em R}}

\def\IN{\relax{\rm I\kern-.15em N}}
\makeatletter
\newdimen\normalarrayskip              % skip between lines
\newdimen\minarrayskip                 % minimal skip between lines
\normalarrayskip\baselineskip
\minarrayskip\jot
\newif\ifold             \oldtrue            \def\new{\oldfalse}
\def\arraymode{\ifold\relax\else\displaystyle\fi} % mode of array entries
\def\@arrayskip{\ifold\baselineskip\z@\lineskip\z@
     \else
     \baselineskip\minarrayskip\lineskip2\minarrayskip\fi}
\def\@arrayclassz{\ifcase \@lastchclass \@acolampacol \or
\@ampacol \or \or \or \@addamp \or
   \@acolampacol \or \@firstampfalse \@acol \fi
\edef\@preamble{\@preamble
  \ifcase \@chnum
     \hfil$\relax\arraymode\@sharp$\hfil
     \or $\relax\arraymode\@sharp$\hfil
     \or \hfil$\relax\arraymode\@sharp$\fi}}
\def\@array[#1]#2{\setbox\@arstrutbox=\hbox{\vrule
     height\arraystretch \ht\strutbox
     depth\arraystretch \dp\strutbox
     width\z@}\@mkpream{#2}\edef\@preamble{\halign \noexpand\@halignto
\bgroup \tabskip\z@ \@arstrut \@preamble \tabskip\z@ \cr}%
\let\@startpbox\@@startpbox \let\@endpbox\@@endpbox
  \if #1t\vtop \else \if#1b\vbox \else \vcenter \fi\fi
  \bgroup \let\par\relax
  \let\@sharp##\let\protect\relax
  \@arrayskip\@preamble}

\begin{document}
\thispagestyle{empty}

\begin{center}
{\huge \bf Chiral Symmetry Breaking on the Lattice: a Study of the Strongly Coupled Lattice Schwinger Model}\footnote{This work 
is supported in part by the Natural
Sciences and Engineering Research Council of Canada, the Istituto
Nazionale di Fisica Nucleare and NATO CRG 930954.}\\
\vskip 0.3 truein
{\bf F. Berruto$^{(a)}$, G. Grignani$^{(a)}$, G. W. Semenoff$^{(b)}$ and P.
Sodano$^{(a)}$}
\vskip 0.3truein
$(a)$ {\it Dipartimento di Fisica and Sezione
I.N.F.N., Universit\`a di Perugia, Via A. Pascoli I-06123 Perugia,
Italy}\\
\vskip 1.truecm
$(b)$ {\it Department of Physics and Astronomy\\University of British 
Columbia\\
6224 Agricultural Road\\Vancouver, British Columbia, Canada V6T 1Z1}\\
\vskip 1.0truein
DFUPG-190-97, UBC/GS-6-97
\vskip 1.0truein
{\bf Abstract}
\vskip 0.3truein
\end{center}
We revisit the strong coupling limit of the Schwinger model on a lattice using 
staggered fermions and the hamiltonian approach to lattice gauge theories.
Although staggered fermions have no continuous chiral symmetry, they posses a discrete axial invariance 
which forbids fermion mass and which must be broken in order for the lattice Schwinger model to exhibit the 
features 
of the spectrum of the continuum theory. We show that this discrete symmetry is indeed broken spontaneously 
in the 
strong coupling limit. 
Expanding around a gauge invariant ground state and carefully considering the normal ordering of the 
charge operator, we derive an 
improved strong coupling expansion and compute
the masses of the low-lying bosonic excitations as well as the 
chiral condensate of the
model. We find very good agreement between our lattice calculations and known continuum values for these quantities already in the fourth order of strong
coupling perturbation theory.  
We also find the exact 
ground state of the antiferromagnetic Ising spin chain with long range 
Coulomb interaction, which determines the nature of the ground state in the strong coupling limit.

\newpage
\setcounter{page}1

\section{Introduction}

Anomalies occur when all of the symmetries of a classical field theory cannot 
be realized simultaneously in the corresponding quantum field theory . The 
classic example is the Adler-Bell-Jackiw anomaly \cite{adler} in a vector-like gauge 
theory where gauge invariance and global axial symmetries are incompatible.  This is a manifestation of the fact that it is impossible to find a 
regularization of the ultraviolet divergences preserving both axial and gauge 
invariance. If gauge invariance is preserved, the observable manifestation of 
the anomaly is the failure of the conservation laws for axial currents and the 
resulting absence of their consequences in the spectrum of the quantized 
theory . 

An interesting issue which arises in the lattice regularization of gauge theories is how the lattice theory produces the effects of the axial anomaly.
Lattice field 
theory is manifestly gauge invariant by construction and therefore, to produce continuum theories, they must find a way to violate axial current conservation.   Normally, in a lattice gauge theories axial anomalies are either cancelled 
by fermion 
doubling or else the lattice regularization breaks the axial symmetry 
explicitly \cite{niels}. The lattice Schwinger model, in the Hamiltonian (latticized space, continuum time) formalism \cite{hwg} with one species of staggered fermions\cite{ssf}\cite{kkg}, represents the unique example where neither of these occur.  The e
ffects of the anomaly are not cancelled by doubling  since the continuum limit of a 1+1-dimensional staggered fermion produces exactly one Dirac fermion 
which is the matter content of the model (see Appendix A). 
Furthermore, even though the continuous axial symmetry is broken 
explicitly by the staggered fermions, a discrete axial 
symmetry remains.   It corresponds to the continuum transformation    
\begin{equation}
\psi(x)\longrightarrow \gamma^{5}\psi(x),\quad\quad 
\overline{\psi}(x)\longrightarrow
-\overline{\psi}(x)\gamma^{5}
\label{symm},
\end{equation}
and appears in the lattice theory as a translation by one site. 
Since the mass operator $\bar\psi\psi$ is odd under
(\ref{symm}), if this symmetry is unbroken then  $<\bar\psi\psi>=0$. 
However, in the continuum theory it is known that \cite{hoso},
\begin{equation}
\left< \bar\psi(x)\psi(x)\right>=-\frac{e^\gamma}{2\pi}\frac{e_c}{\sqrt{\pi}}
~~.
\label{chico}
\end{equation}
where $\gamma=0.577...$ is the Euler constant.
The expectation value in eq.(\ref{chico}) is the chiral condensate.
\footnote{It is worth to stress that the sign of the chiral condensate in eq.(\ref{chico}) is just a convention. The chiral condensate has always the opposite sign of the fermion mass $m$ that appears in the Lagrangian of the theory: it is the order param
eter for a spontaneous symmetry breaking of the chiral symmetry. The chiral condensate is the analogous of the magnetization for a spin model and the mass m plays the role of the magnetic field. So even if one is studying a massless model, in order to dec
ide the sign of the chiral condensate, one has to consider a small mass term $m$ that then one sends to zero. The minus sign which appears in eq.(\ref{chico}) is due to the fact that one considers the $m\longrightarrow 0^{+}$ limit in the Schwinger model 
Lagrangian.} 
This non-zero value of the chiral condensate 
is one of  the main effects of the anomaly. 
In order for the lattice theory to produce this and other features 
of the continuum theory, the discrete axial symmetry (\ref{symm}) 
must be spontaneously broken. 

The continuum Schwinger model \cite{scw,low} 
 is the prototypical example of a solvable model where the anomaly occurs.
It is 1+1-dimensional electrodynamics with a single charged massless Dirac spinor field.  The action is 
\begin{equation}
S = \int d^{2} x (\overline{\psi}(i\gamma_{\mu}\partial^{\mu}+\gamma_{\mu}
A^{\mu})\psi-\frac{1}{4e^{2}_{c}}F_{\mu\nu}F^{\mu\nu})
\end{equation}
It is invariant under gauge transformations,
\begin{equation}
\psi(x)\longrightarrow e^{i\theta(x)}\psi(x),\psi^{\dag}(x)\longrightarrow \psi^{\dag}(x)
e^{-i\theta(x)}
\end{equation}
\begin{equation}
A_{\mu}\longrightarrow A_{\mu}+\partial_{\mu}\theta(x)
\end{equation}
and, formally, under the axial phase rotation,  
\begin{equation}
\psi(x)\longrightarrow e^{i\gamma_{5} \alpha} \psi(x),
\psi^{\dag}(x)\longrightarrow \psi^{\dag}(x) e^{-i\gamma_{5} \alpha}
\end{equation}
(where $\gamma^{5}=i\gamma^{0}\gamma^{1}$). 

At the classical level the above 
symmetries lead to conservation laws for the vector and axial currents, 
\begin{equation}
j^{\mu}(x)=\overline{\psi}(x)\gamma^{\mu}\psi(x),\quad\quad j_{5}^{\mu}(x)
=\overline{\psi}(x)\gamma^{5}\gamma^{\mu}\psi(x),
\end{equation}
respectively. At the quantum level both currents cannot be simultaneously 
conserved. If the regularization is gauge invariant, so that the vector current 
is conserved, then the axial current obeys the anomaly equation \cite{cole} 
\begin{equation}
\partial_{\mu} j_{5}^{\mu}(x)=\frac{e^{2}_{c}}{2\pi}\epsilon^{\mu \nu}F_{\mu\nu}(x).
\end{equation}
Moreover, the correlation functions of the model do not 
exhibit axial symmetry.

There are only neutral particles in the spectrum of the Schwinger model; notably, the bound state boson 
created by the axial current operator, 
\begin{equation}
\partial^{\mu}\phi(x)=\sqrt{\pi}j_{5}^{\mu}(x)
\label{anomaly}
\end{equation}
appears in the spectrum as a free pseudoscalar with mass $\frac{e_{c}}{\sqrt{\pi}}$.

In terms of the boson field, the action is 
\begin{equation}
S=\int d^{2}x(\frac{1}{2}\partial_{\mu}\phi
\partial^{\mu}\phi+\frac{e^{2}_{c}}{2\pi}\phi^{2})
\end{equation}
and the correlation functions for the vector and axial currents are obtained 
from the correlation functions of derivatives of the bose fields. 
The coupling constant $e_{c}$ has the dimension of a mass and the model
is super-renormalizable. 

Since the Schwinger model is solvable it has often been used as a field theory 
where methods of lattice gauge theory, particularly the strong coupling expansion can be tested and compared with exact results of the continuum model \cite{kog}\cite{sus}. 

In this Paper, we shall revisit the lattice quantization of the Schwinger 
model.  Our main purpose is to demonstrate two results.  One is to show how the effects of the anomaly appear through spontaneous symmetry breaking.  
The other is to 
demonstrate that a strong coupling expansion around a gauge, parity and charge
conjugation invariant lattice ground state can produce results in 
good agreement with the continuum.   Our results turn out to be quite 
different from those of previous strong coupling computations \cite{kog}\cite{sus}.  The difference arises from our careful treatment of the normal 
ordering of the charge density operator, which must be defined so as to 
be compatible with all of the discrete symmetries of the theory.  The 
necessity of careful normal ordering of the charge operator was pointed 
out in \cite{dssl}.  
Our vacuum state, vacuum energy and energies of elementary excitations are
different from those in \cite{kog,sus,ham} and, as we shall show, they give a more accurate extrapolation to the continuum limit.

The Hamiltonian and gauge constraint of the continuum Schwinger model are
\begin{equation}
H=\int dx\left( \frac{e_c^2}{2}E^2(x)+
\psi^{\dagger}(x)\alpha\left(i\partial_x +eA(x)\right)\psi(x)\right)
\end{equation}
\begin{equation}
\partial_x E(x)+\psi^{\dagger}(x)\psi(x)\sim 0
\end{equation}
A lattice Hamiltonian and constraint which reduce to these in the continuum limit are:  
\begin{equation}
H_{S}=\frac{e^{2}_{L}a}{2}\sum_{x}E_{x}^{2}-\frac{it}{2a}\sum_{x}(\psi_{x+1}
^{\dag}e^{iA_{x}}\psi_{x}-\psi_{x}^{\dag}e^{-iA_{x}}\psi_{x+1})
\label{hamilton}
\end{equation}
\begin{equation}
E_{x}-E_{x-1}+\psi_{x}^{\dag}\psi_{x}-\frac{1}{2}\sim 0\ ,
\label{gauss}
\end{equation}
where the fermion fields are defined on the sites, $x=-\frac{N}{2},
-\frac{N}{2}+1,...,\frac{N}{2}$, gauge and the electric fields, $ A_{ x}$ and
 $E_{x}$,  on the links $[x; x + 1]$, $N$ is an even integer 
and, when $N$ is finite, we use periodic boundary conditions.  When $N$ is finite, the continuum limit is the Schwinger model on a circle \cite{man,hoso}.
The coefficient $t$ of the hopping term in (\ref{hamilton})
plays the role of the lattice light speed. In the naive continuum limit,
$e_L=e_c$ and $t=1$.  However, we shall keep $e_L$ and $t$ as parameters 
which can be adjusted to fit the lattice values of quantities such as the 
mass gap and  the chiral condensate to those which are known in the 
continuum.

The non vanishing (anti-)commutators for the lattice variables are
\begin{equation}
[A_{x},E_{y}]=i\delta_{xy},\quad\quad\{\psi_{x},\psi_{y}^{\dag}\}=\delta_{xy}
\end{equation}

The Hamiltonian and gauge constraint exhibit the discrete symmetries 
\begin{itemize}
\item{}Parity P: 
\begin{equation}
A_{x}\longrightarrow -A_{-x-1},\ E_{x}\longrightarrow -E_{-x-1},\ 
\psi_{x}\longrightarrow (-1)^{x}\psi_{-x},\ \psi_{x}^{\dag}\longrightarrow 
(-1)^{x}\psi_{-x}^{\dag}
\label{par}
\end{equation}

\item{}Discrete axial symmetry $\Gamma$: 
\begin{equation}
A_{x}\longrightarrow A_{x+1},\ E_{x}\longrightarrow E_{x+1},\ 
\psi_{x}\longrightarrow \psi_{x+1},\ \psi_{x}^{\dag}\longrightarrow 
\psi_{x+1}^{\dag}
\label{chir}
\end{equation}

\item{}Charge conjugation C:
\begin{equation}
A_{x}\longrightarrow -A_{x+1},\  E_{x}\longrightarrow -E_{x+1},\ 
\psi_{x}\longrightarrow \psi^{\dag}_{x+1},\ \psi_{x}^{\dag}\longrightarrow 
\psi_{x+1}
\label{char}
\end{equation}
When $A_x=0$, the spectrum of the hopping Hamiltonian is $\epsilon(pa)= 
t \sin pa (p\in 
[0,\frac{2 \pi}{a} ])$. In the low energy limit it resembles a massless 
relativistic spectrum for excitations near the Fermi level. 
The two 
intersections of the energy band with the Fermi level provide the continuum 
right and left moving massless fermions. The electron ground state is invariant under charge conjugation only when the Fermi level 
is at $\epsilon(p_F)=0$ (i.e. when exactly half of the fermion states are 
filled and $\sum_x<\rho(x)>=0$).  In the remainder of this paper we consider only this case of half-filling.
\end{itemize}

One has two possile ways to put the spinors  $\psi=\left(\begin{array}{c}
\psi_{1}\\
\psi_{2}
\end{array}\right)$ on the lattice in the staggered fermion formalism. One can put upper components on even sites and lower components on odd sites or viceversa.
The mass operator, which reduces to $\bar\psi\psi$ in the continuum limit is
the staggered charge density
\begin{equation}
M(x)=\frac{(-1)^x}{a}\left(\psi_x^{\dagger}\psi_x-1/2\right)
\label{sma}
\end{equation}
for the first choice of staggered fermions or
\begin{equation}
M(x)=\frac{(-1)^{x+1}}{a}\left(\psi_x^{\dagger}\psi_x-1/2\right)
\label{smbb}
\end{equation}
in the other case.

These operators (\ref{sma})(\ref{smbb}) are scalars under parity and are even under
charge conjugation but change sign under a discrete axial transformation.

The lattice Schwinger model is equivalent to a one 
dimensional quantum Coulomb gas on the lattice. To see this one can fix the 
gauge, $A_{x} = A$ (Coulomb gauge). Eliminating the non-constant electric 
field and using the gauge constraint, one obtains the effective Hamiltonian
\begin{eqnarray}
H_{S}&=&H_u+H_p
\equiv\left[\frac{e^{2}_{L}}{2 N}E^{2}+\frac{e^{2}_{L}a}{2}
\sum_{x,y}\rho(x) V(x-y)\rho(y)\right]+\nonumber\\
&+&\left[
-\frac{it}{2a}\sum_{x}(\psi_{x+1}^{\dag}e^{iA}\psi_{x}-\psi_{x}^{\dag}e^{-iA}
\psi_{x+1})\right]\ ,
\label{hs}
\end{eqnarray}
where the charge density is
\begin{equation}
\rho(x)=\psi^{\dag}_x\psi_x-\frac{1}{2}\ \ ,
\end{equation}
and the potential
\begin{equation}
V(x-y)=\frac{1}{N}
\sum^{N-1}_{n=1} e^{i 2\pi n (x-y)/N}\frac{1}{4\sin^2\frac{\pi n}{N}}
\end{equation}
is the Fourier transform of the inverse laplacian on the lattice
for non zero momentum.
The constant electric field is normalized so that $[ A, E ] = i$ . 
The constant 
modes of the gauge field decouple in the thermodynamic limit 
$ N \longrightarrow \infty $.
The gauge fixed Hamiltonian (\ref{hs}) can also be written as a quantum spin model.  Consider the Jordan-Wigner transformation
\begin{equation}
\psi_x=\prod_{y<x}[2iS_3(y)]S^-(x)~~,~\psi^{\dagger}_x=
\prod_{y<x}[-2iS_3(y)]S^+(x)~~,~\psi^{\dagger}_x\psi_x=S^3(x)+1/2
\end{equation}
Then, the Hamiltonian is 
\begin{equation}
H=\frac{t\cos(A)}{a}\sum_x\left( S^1(x)S^1(x+1)+S^2(x)S^2(x+1)\right)+
\frac{e_L^2a}{2}\sum_{x,y}S^3(x)V(x-y)S^3(y)+e_L^2E^2/2N
\label{spinham}
\end{equation}
For the moment, we ignore the constant modes $E$ and $A$. 
The first term in the Hamiltonian 
is the quantum x-y model which has a disordered 
ground state (of course, exact solution of this model is obtained by 
the inverse Jordan-Wigner transformation \cite{lsm,nambu}).  
The second is a long-ranged Ising interaction.  
We will prove (see also ref. \cite{dssl}) 
in the next section that, in spite of the low 
dimensionality of this system, the latter term has a N\'eel ordered 
ground state.  This is a result of the infinite range of the Coulomb interaction.  
When both terms are present in the Hamiltonian, they compete, the 
first one favoring disorder and the second one favoring order.  Later, 
when we extrapolate our strong coupling results to the continuum limit, 
we shall assume that the ordered ground state persists for all positive 
values of the dimensionless constant $e_L^2a^2/t$.  The strong coupling 
limit is where this constant is large and the ground state has N\'eel order.  
The continuum limit of the Schwinger model is found where this constant is
 small.  We shall assume that the N\'eel order persists in that limit.  
\footnote{If it did not persist, but there was a second order phase 
transition at some critical value of the constant to a disordered 
state, the continuum limit would be found at that phase transition.}

There are two interesting questions which we shall not pursue in this paper, 
but will consider later.  One is the possibility of finding exact solutions 
of (\ref{spinham}) with $E$ and $A$ set to zero using the Bethe ansatz.  This would effectively give an exact solution of the lattice Schwinger model in
infinite volume.  The
other is the solution of the model (\ref{spinham}) when the topological modes,
$E$ and $A$, are 
included but the Coulomb interaction is ignored.  The solution of this model would give some insight into how the anomaly arises in the weak coupling limit of the lattice theory.  

In section 2 we set up the strong coupling expansion around the 
gauge invariant ground state of the 
unperturbed Hamiltonian $H_u$. Then, we compute to the fourth order in powers of $\epsilon=\frac{t}{2e_{L}^{2}a^{2}}$,
the corrections to the energies of the ground state and of the low-lying
bosonic excitations (at zero momentum). 
Subtracting the energy of the ground state from those of the low-lying excitations defines the 
lattice meson masses. 
In section 3 we compute the chiral 
condensate to the same order in $\epsilon$. 
Section 4 is devoted to the comparison of 
the lattice results with the continuum ones; there we shall see that, having used a gauge 
invariant ground state, 
the improved strong coupling  expansion leads to a by far more accurate evaluation of the 
physical parameters then the strong coupling expansions given 
in \cite{kog,sus,ham}. 
Section 5 shortly comments on our results.

In the appendices we provide some technical details.

\section{The low-lying excitation spectrum}

\setcounter{equation}{0}
We shall examine the Schwinger model in the strong coupling limit. 
To solve the lattice Schwinger model in the strong coupling expansion, it is 
necessary to find states  which are annihilated by the generator of gauge transformations
\begin{equation}
E_{x}-E_{x-1}+\psi^{\dag}_{x}\psi_{x}-\frac{1}{2}
\label{gauss}
\end{equation}
and, at the same time, are eigenstates of the unperturbed Hamiltonian 
\begin{equation}
H_{u}=\frac{e_{L}^{2}}{2}\sum_{x}E_{x}^{2}\ .
\end{equation}
We shall quantize the gauge fields in the functional Schr\"{o}dinger picture where the wave-functions are functions of the 
variables $A_{x}$ and the electric field operators are defined as  
\begin{equation}
E_{x}=\frac{1}{i}\frac{\delta}{\delta A_{x}}\ .
\end{equation}
In a previous publication \cite{dssl} it was proved that there were two gauge invariant ground states of the system. In fact, when the number, N, 
of lattice sites is even the two states   
\begin{eqnarray}
|\psi>&=&\prod_{x=even}\psi_{x}^{\dag}|0>e^{-i\sum_{x}A_{x}\frac{(-1)^{x}}{4}}\\
|\chi>&=&\prod_{x=odd}\psi_{x}^{\dag}|0>e^{i\sum_{x}A_{x}\frac{(-1)^{x}}{4}}
\end{eqnarray}
are degenerate gauge invariant ground states of $H_u$.
To see this, note that then $H_u$ acts on gauge invariant states, it is effectively the sum of two operators, the zero mode energy $e_L^2 E^2/2N$ 
where $E=\sum_x E_x/\sqrt{N}$ and the Coulomb energy
\begin{eqnarray}
\frac{e^{2}_{L}a}{2}
\sum_{x,y}\rho(x) V(x-y)\rho(y)=\frac{e^{2}_{L}a}{2N}\sum_{k\neq
0}\rho(-k)\frac{1}{4\sin^{2}\frac{k}{2}}\rho(k)\ .
\label{huuu}
\end{eqnarray}
where we have used the Fourier transform
\begin{equation}
\psi^{\dag}_{x}\psi_{x}-\frac{1}{2}=
\frac{1}{\sqrt{N}}\sum_{p}e^{ipx}\rho(p)\ ,\ p=\frac{2\pi n}{N}\ ,\ n\in \{-\frac{N}{2},...,\frac{N}{2}-1\}
\end{equation}
Furthermore, gauge invariant states have vanishing total charge,
\begin{equation}
\sum_{x}(\psi_{x}^{\dag}\psi_{x}-\frac{1}{2})=0
\end{equation}
The Coulomb energy satisfies the bound
\begin{eqnarray}
\frac{e^{2}_{L}a}{2N}\sum_{k\neq 0}\rho(-k)\frac{1}{4\sin^{2}\frac{k}{2}}\rho(k)\geq \frac{e^{2}_{L}a}{2N}\sum_{k\neq 0}\rho(-k)\frac{1}{4}\rho(k)=\nonumber\\
=\frac{e^{2}_{L}a}{8}\sum_{x}(\psi_{x}^{\dag}\psi_{x}-\frac{1}{2})^{2}
=e_L^2a/32 
\end{eqnarray} 
(since  
$(\psi_{x}^{\dagger}\psi_{x}-\frac{1}{2})^{2}=\frac{1}{4}$). 
The states $|\psi>$ and $|\chi>$ are eigenstates of the charge density 
with eigenvalues
$\rho(k)=\pm\frac{1}{2}\delta{k,\pi}$.  These charge densities  
saturate the lower bound for the Coulomb energy.  Furthermore, $E|\psi>=0=E|\chi>$ so $|\psi>$ and $|\chi>$ minimize the energy of the zero mode.  Therefore, they are degenerate ground 
states of $H_u$.

The ground states
\label{smb} are invariant under parity and charge conjugation but they 
both break chiral symmetry, i.e. the symmetry under translations by one site.  
The order parameter of the chiral symmetry, the mass operator $M=\frac{1}{Na}
\sum_xM(x)$, is diagonal in these states and has 
eigenvalue $\frac{1}{2}$. 
\footnote{One has to consider $M(x)$ given by eq.(\ref{sma}) when applying it to $|\psi>$ and $M(x)$ given by eq.(\ref{smbb}) when applying it to $|\chi>$.}
In order to have a chiral condensate of negative sign as in the continuum, one has to consider the mass operator $M'(x)=-M(x)$. 
It is interesting that 
the ground states are degenerate and are not invariant under chiral 
symmetry, even in finite volume and in 
one spatial dimension ; it is possible due to 
the long-ranged nature of the Coulomb interaction\cite{cteo}. The chiral symmetry is spontaneously broken in the infinite volume limit.
It is essential that the number of sites of the lattice, N, is even. 

Of course once perturbations from the hopping term in the fermion Hamiltonian
are taken into account, when the volume is finite, there are high order terms 
(of order $N$) in the strong coupling perturbation theory which mix the 
ground states that are constructed by perturbing $|\psi>$ and $|\chi>$.
The mixing amplitude is
\begin{equation}
<\chi|H_{h}(\frac{1}{E_{0}-H_{C}}H_{h})^{N-1}|\psi>=-\frac{N^{2}}{2}
\frac{|t|^{N}}{2^{N}}\frac{4^{N-1}}{e_{L}^{2(N-1)}}
\frac{N!}{((\frac{N}{2})!)^{2}} 
\label{mixa}
\end{equation}
Since the amplitude (\ref{mixa}) is negative and real , whatever the diagonal elements are, the state with lower energy is the state
\beq
\frac{1}{\sqrt{2}}(|\psi>+|\chi>)
\eeq
This is an eigenstate of the discrete axial transformation, with eigenvalue 
$1$. Thus the true ground state is chirally symmetric. 

In order to avoid 
this restoration of the symmetry, it is necessary to take the infinite  
N limit so that the two ground states are not mixed to any finite order of 
perturbation theory. Thus, in the perturbative expansion one only has to consider diagonal matrix elements and, consequently, perturbation theory 
for non-degenerate-states. We shall then choose as unperturbed ground 
state $|\psi>$. The results concerning the energies would have been the same 
had we chosen 
$|\chi>$.

Excitations are created by the operators
\begin{equation}
\psi_{j}^{\dag}e^{i\sum_{x=i}^{j}A_{x}}\psi_{i}
\end{equation}
and have energy $\frac{e_{L}^{2}}{2}|i-j|$ greater than the ground state energy. A single electron, in order to have a gauge invariant state, 
must be attached to a line of electric flux which goes from the electron to infinity. This is not allowed at all in a finite volume with 
periodic boundary conditions and on the infinite lattice, the string of electric flux would have infinite energy. Thus electric charges are 
confined and the string tension is given by the electric charge $\frac{e_{L}^{2}}{2}$.

Let us now start the perturbative computation of the excitations mass gap.
We shall work in the Coulomb gauge $\nabla A=0$, so that $A$ is $x$-independent 
and the ground states are
\begin{eqnarray}
|\psi>&=&\prod_{x=even}\psi_{x}^{\dag}|0>\\
|\chi>&=&\prod_{x=odd}\psi_{x}^{\dag}|0>.
\end{eqnarray}   
The Coulomb potential reads 
\begin{equation}
V(x-y)=\frac{1}{N}\sum_{k\neq 0} \frac{1}{4\sin^{2}\frac{k}{2}}e^{ik(x-y)
}=\frac{1}{2N}(x-y)^{2}-\frac{1}{2}|x-y|+\frac{1}{12N}(N^{2}-1).
\end{equation}
The Schwinger Hamiltonian, rescaled by the factor $e_{L}^{2}a$,
 may then be written as
\begin{equation}
H=H_{0}+\epsilon H_{h}=\frac{1}{e_{L}^{2}a}(H_{C}+H_{p})
\end{equation}
with 
\begin{equation}
H_{0}=\sum_{x>y}[\frac{(x-y)^{2}}{2N}-\frac{(x-y)}{2}]\rho(x)\rho(y)
\end{equation}
\begin{equation}
H_{h}=-i\sum_{x}(\psi_{x+1}^{\dag}e^{iA}\psi_{x}-\psi_{x}^{\dag}e^{-iA}
\psi_{x+1})=-i(H_{R}-H_{L})
\end{equation}
and
\begin{equation}
\epsilon=\frac{t}{2e^{2}_{L}a^{2}}.
\end{equation}
To the fourth order in $\epsilon$ the perturbative expansion for the energy is 
given by\footnote{
We observe that for the ground state 
perturbative expansion the relative minus sign in the hopping 
Hamiltonian is irrelevant whereas it will play an important role in the
calculation of the excited state energy. 
This is easily seen if one notices that at the second order in $\epsilon$  
one gets
\begin{eqnarray}
E_{\psi}^{(2)}&=&<\psi|(H_{L}-H_{R})\frac{\Pi_{\psi}}{E_{\psi}^{(0)}-H_{0}}
(H_{R}-H_{L})|\psi>=\nonumber\\
&=&<\psi|H_{L}\frac{\Pi_{\psi}}{E_{\psi}^{(0)}-H_{0}} 
H_{R}|\psi>+<\psi|H_{R}\frac{\Pi_{\psi}}{E_{\psi}^{(0)}-H_{0}}H_{L}|\psi>\nonumber
\end{eqnarray}}
\begin{equation}
E_{\psi}=E_{\psi}^{(0)}+\epsilon^{2}E_{\psi}^{(2)}+\epsilon^{4}E_{\psi}^{(4)}
\end{equation}
where
\begin{equation}
E_{\psi}^{(0)}=<\psi|H_{0}|\psi>
\end{equation}
\begin{equation}
E_{\psi}^{(2)}=<\psi|H_{h}^{\dag}\frac{\Pi_{\psi}}{E_{\psi}^{(0)}-H_{0}}H_{h}|\psi>
\end{equation}
\begin{eqnarray}
E_{\psi}^{(4)}=<\psi|H_{h}^{\dag}\frac{\Pi_{\psi}}{E_{\psi}^{(0)}-H_{0}}H_{h}^{\dag}
\frac{\Pi_{\psi}}{E_{\psi}^{(0)}-H_{0}}H_{h}\frac{\Pi_{\psi}}{E_{\psi}^{(0)}-H_{0}}H_{h}
|\psi>+\nonumber\\
-<\psi|H_{h}^{\dag}\frac{\Pi_{\psi}}{E_{\psi}^{(0)}-H_{0}}H_{h}|\psi>
<\psi|H_{h}^{\dag}\frac{\Pi_{\psi}}{(E_{\psi}^{(0)}-H_{0})^{2}}H_{h}|\psi>
\end{eqnarray}
and
$1-\Pi_{\psi}$ is a projection operator 
which projects a generic state $|s>$ on $|\psi>$
\begin{equation}
(1-\Pi_{\psi})|s>=|\psi>\ .
\end{equation}

The phases $e^{iA}$ in the hopping Hamiltonian are irrelevant 
until we reach the $N^{th}$ order in the perturbative expansion. 
As a matter of fact, 
only at this order the transverse electric field becomes important. 
By direct evaluation one gets  
\begin{eqnarray}
E_{\psi}^{(0)}=\frac{N}{32}\\
E_{\psi}^{(2)}=-4N\\
E_{\psi}^{(4)}=192N
\end{eqnarray}
\begin{equation}
E_{\psi}=\frac{N}{32}-4N\epsilon^{2}+192N\epsilon^{4}.
\end{equation}

Next we compute the energies of the low-lying excitations
to the same order in the perturbative expansion.
The lowest -lying excitations are degenerate and are created by the operators 
\begin{eqnarray}
H_{L}(x)&=&\psi_{x}^{\dag}e^{iA}\psi_{x+1}\\
H_{R}(x)&=&\psi_{x+1}^{\dag}e^{-iA}\psi_{x}
\end{eqnarray}
which move a particle one lattice spacing to the left and right respectively. When they act on the vacuum $|\psi>$ they create the two states 
\begin{eqnarray}
|x,R>&\equiv&H_{R}(x)|\psi>=\psi_{x+1}^{\dag}e^{-iA}\psi_{x}|\psi>\\
|x,L>&\equiv&H_{L}(x)|\psi>=\psi_{x}^{\dag}e^{iA}\psi_{x+1}|\psi>
\end{eqnarray}
which are degenerate in energy (see eq.(\ref{bba}) in appendix B).
In order to resolve the degeneracy of these two states , one must find the first nontrivial order of the 
perturbative expansion in which the two states are mixed; this turns out to be the 
second order. It is easy to show that the second order of perturbative expansion is diagonal  in the two states which are created  by the parity odd and 
even operators
\begin{eqnarray}
H_{-}(x)&\equiv&H_{R}(x)+H_{L}(x)\\ 
H_{+}(x)&\equiv&H_{R}(x)-H_{L}(x)\ .
\end{eqnarray}
In fact
\begin{equation}
P H_{\pm}(x)=\pm H_{\pm}(-x)\ .
\end{equation}
The parity odd operator creates the state  with lower energy. Thus, it is this the state which 
will have finite energy gap in the continuum limit 
and which will correspond to the pseudoscalar boson of the continuum Schwinger model.
In the continuum one gets this massive excitation by 
applying the vector flux $j^{1}=j_{5}^{0}$ to the vacuum. The excitation created by $H_{+}$ 
is the scalar excitation which in the continuum limit decouples from the spectrum. 
Since one is interested in the masses of these excitations, one has to compute their energies at zero momentum.

On the lattice the construction of the pseudoscalar excitation at zero momentum is 
provided by
\begin{eqnarray}
|\theta>&\equiv&\frac{1}{\sqrt{N}}\sum_{x=1}^N|\theta(x)>=\frac{1}{\sqrt{N}}\sum_{x=1}^{N}H_{-}(x)|\psi>=
\frac{1}{\sqrt{N}}\sum_{x=1}^{N}(\psi_{x}^{\dag}e^{iA}\psi_{x+1}+
\psi^{\dag}_{x+1}e^{-iA}\psi_x)|\psi>=\nonumber\\
&=&\frac{1}{\sqrt{N}}\sum_{x=1}^{N}j^{1}(x)|\psi>=
\frac{1}{\sqrt{N}}\sum_{x=1}^{N}
j_{5}^{0}(x)|\psi>.
\end{eqnarray}
One could also consider the scalar massive excitation 
\begin{eqnarray}
|\varphi>&\equiv&=\frac{1}{\sqrt{N}}\sum_{x=1}^{N}|\varphi(x)>=\frac{1}{\sqrt{N}}\sum_{x=1}^{N}H_{+}(x)|\psi>=\nonumber\\
&=&
\frac{1}{\sqrt{N}}\sum_{x=1}^{N}(\psi_{x}^{\dag}e^{iA}\psi_{x+1}-
\psi^{\dag}_{x+1}e^{-iA}\psi_x)|\psi>=\frac{1}{\sqrt{N}}\sum_{x=1}^{N}j^{5}(x)|\psi>;
\label{pseudo}
\end{eqnarray}
The states $|\theta>$ and $|\varphi>$ are characterized 
by a dimer-antidimer configuration; that is by a pair of particles 
followed or preceded by a pair of antiparticles.

The energy of the state $|\theta>$, 
at the second order in $\epsilon$, is given by 
\begin{equation}
E_{\theta}^{(2)}=\sum_{x,y}\frac{1}{N}<\theta(y)|(H_{L}-H_{R})
\Lambda_{\theta}(H_{R}-H_{L})|\theta(x)>
\end{equation}
where
\begin{equation}
\Lambda_{\theta}=\frac{\Pi_{\theta}}{E_{\theta}^{(0)}-H_{0}}
\end{equation}
with $E_{\theta}^{(0)}=E_{\psi}^{(0)}+\Delta$, $\Delta=\frac{1}{4}$ and $1-\Pi_{\theta}$ a projection 
operator onto $|\theta>$.
A state obtained by applying the hopping Hamiltonian to the state $|\theta>$ 
has a dimer-antidimer pair; as a consequence its energy will be
raised of $2\Delta$ with respect to $E_{\psi}^{(0)}$. 
\footnote{One must be careful in applying one more time the hopping Hamiltonian.
 Since the
Coulomb Hamiltonian and the hopping Hamiltonian do not commute (see appendix B), 
there is a state created 
by appling two times the hopping Hamiltonian to the state $|\theta>$ which has not a gap of $3\Delta$ 
 with respect to the ground
state, but it has a gap of $5\Delta$-characterized by three particles 
followed or preceded by three antiparticles,
that is, it is created by applying $H_{L}^{3}$ or $H_{R}^{3}$ to the ground state .}
The computation of $E_{\theta}^{(2)}$ (see appendix C) leads to
\begin{equation}
E_{\theta}^{(2)}=-
\frac{1}{\Delta}N+\frac{2}{\Delta}=-4N+8
\end{equation}
and 
\begin{equation}
<\theta|H_{h}^{\dag}\Lambda_{\theta}^{2}
H_{h}|\theta>=\frac{N}{\Delta^{2}}-\frac{2}{\Delta^{2}}=16N-32\ .
\end{equation}

Next one should compute 
\begin{eqnarray}
E_{\theta}^{(4)}=\frac{1}{N}\sum_{x,y}<\theta(y)|H_{h}^{\dag}
\Lambda_{\theta}H_{h}^{\dag}\Lambda_{\theta}H_{h}\Lambda_{\theta}H_{h}|\theta(x)>  
+\nonumber\\
-\frac{1}{N^{2}}\sum_{x,y}<\theta(y)|H_{h}^{\dag}\Lambda_{\theta}H_{h}
|\theta(x)>\sum_{x,y}<\theta(y)|H_{h}^{\dag}\Lambda_{\theta}^{2}H_{h}|\theta(x)>
\label{et4}.
\end{eqnarray}
Using the matrix elements given in appendix C, one gets
\begin{equation}
E_{\theta}^{(4)}=\frac{3}{\Delta^{3}}N-\frac{9}{\Delta^{3}}=192N-576\ .
\end{equation}
The pseudoscalar excitation energy, to the fourth order in $\epsilon$, 
is then given by
\begin{equation}
E_{\theta}=(\frac{N}{32}+\frac{1}{4})+(-4N+8)\epsilon^{2}+(192N-576)
\epsilon^{4}\ .
\end{equation}

Using again the matrix element given in appendix C, one easily obtains the 
energy $E_{\varphi}$ of the state $|\varphi>$, $i.e.$
\beq
E_{\varphi}=(\frac{N}{32}+\frac{1}{4})+(-4N+24)\epsilon^{2}+(192N-1600)
\epsilon^{4}\ .
\end{equation}

Taking the difference between $E_{\theta}$ ($E_{\varphi}$) 
and $E_{\psi}$, one gets the masses of the  pseudoscalar and scalar 
excitations , 
\beq
\frac{m^P}{e^{2}_{L}a}=\frac{1}{4}+8\epsilon^{2}-576\epsilon^{4}\ ,
\label{mga}
\end{equation}
\begin{equation}
\frac{m^S}{e^{2}_{L}a}=\frac{1}{4}+24\epsilon^{2}-1600\epsilon^{4}\ .
\label{mgb}
\end{equation}

Eqs.(\ref{mga},\ref{mgb}) provide the behavior of $m^P$ and $m^S$ 
for small values of
$z=\epsilon^{2}=\frac{t^{2}}{4e^{4}_{L}a^{4}}$.

As expected from the continuum theory, 
the pseudoscalar particle mass is smaller 
than the scalar particle mass.
Furthermore, since both quantities are intensive, 
there is no $N$ dependence in 
(\ref{mga},\ref{mgb}). 

Now we compute the dispersion relation of the pseudoscalar and scalar excitations to the fourth order in perturbation theory.

At the zero-th order one has no 
momentum dependence in the lattice pseudoscalar excitation since
\begin{equation}
E_{\theta}^{(0)}(p)=\frac{1}{N}\sum_{x,y}<\theta(y)|H_{0}|\theta(x)>e^{ip(x-y)}=\frac{N}{32}+\Delta=\frac{N}{32}+\frac{1}{4}\ .
\label{eth1}
\end{equation}
The momentum dependence starts only at the second order in $\epsilon$, where one gets 
\begin{equation}
E_{\theta}^{(2)}(p)=\frac{1}{N}\sum_{x,y}<\theta(y)|H_{h}^{\dag}\Lambda_{\theta}H_{h}|\theta(x)>e^{ip(x-y)}\ .
\label{eth2}
\end{equation}
Using the matrix elements given in appendix C, one has
\begin{equation}
E_{\theta}^{(2)}(p)=-\frac{1}{\Delta}(N-4)-\frac{2}{\Delta}\cos pa=-4N+16-8\cos pa\ .
\end{equation}
The dispersion relation, at the fourth order in perturbation theory, reads
\begin{eqnarray}
E_{\theta}^{(4)}(p)=\frac{1}{N}\sum_{x,y}<\theta(y)|H_{h}^{\dag}\Lambda_{\theta}H_{h}^{\dag}\Lambda_{\theta}H_{h}\Lambda_{\theta}H_{h}|\theta(x)>e^{ip(x-y)
}+\nonumber\\
-\frac{1}{N^{2}}\sum_{x,y}<\theta(y)|H_{h}^{\dag}\Lambda_{\theta}H_{h}|\theta(x)>e^{ip(x-y)}\sum_{x,y}<\theta(y)|H_{h}^{\dag}\Lambda_{\theta}^{2}H_{h}|\theta(x)>e^{ip(x-y)}
\end{eqnarray}
and leads to
\begin{eqnarray}
E_{\theta}^{(4)}(p)&=&\frac{3}{\Delta^{3}}N-\frac{1}{2\Delta^{3}}\cos 2pa+\frac{4}{\Delta^{3}}(2\cos pa-\cos^{2} pa)
-\frac{25}{2}\frac{1}{\Delta^{3}}=\nonumber\\
&=&192N-32\cos 2pa+256(2\cos pa-\cos^{2}pa)-800\ .
\label{eth4}
\end{eqnarray}
Using (\ref{eth1}), (\ref{eth2}) and (\ref{eth4}), the dispersion relation of the lattice pseudoscalar excitation is given by 
\begin{eqnarray}
E_{\theta}(p)&=&E_{\theta}^{(0)}(p)+\epsilon^{2}E_{\theta}^{(2)}(p)+\epsilon^{4}E_{\theta}^{(4)}(p)+...=\nonumber\\
&=&E_{\theta}(0)+\frac{2}{\Delta}(1-\cos pa)\epsilon^{2}+[\frac{1}{2\Delta^3}(1-\cos 2pa)-\frac{4}{\Delta^{3}}(1-\cos pa)^{2}]
\epsilon^{4}=\nonumber\\
&=&E_{\theta}(0)+8(1-\cos pa)\epsilon^{2}+[32(1-\cos 2pa)-256(1-\cos pa)^{2}]\epsilon^{4}.
\end{eqnarray}
Performing a small momentum expansion, one gets 
\begin{equation}
E_{\theta}(p)=E_{\theta}(0)+(4\epsilon^{2}+64\epsilon^{4})(pa)^{2}+\dots
\label{mg2}
\end{equation}

The same procedure yields 
\begin{eqnarray}
E_{\varphi}(p)&=&E_{\varphi}^{(0)}(p)+\epsilon^{2}E_{\varphi}^{(2)}(p)+\epsilon^{4}E_{\varphi}^{(4)}(p)+...=\nonumber\\
&=&E_{\theta}(0)-\frac{2}{\Delta}(1-\cos pa)\epsilon^{2}+[\frac{1}{2\Delta^3}(1-\cos 2pa)+\frac{16}{\Delta^{3}}-\frac{4}{\Delta^{3}}(1+\cos pa)^{2}]
\epsilon^{4}=\nonumber\\
&=&E_{\theta}(0)-8(1-\cos pa)\epsilon^{2}+[32(1-\cos 2pa)+1024-256(1+\cos pa)^{2}]\epsilon^{4}.
\end{eqnarray}
which, for small momentum gives
\begin{equation}
E_{\varphi}(p)=E_{\varphi}(0)+(-4\epsilon^{2}+72\epsilon^{4})(pa)^{2}+\dots
\label{mg3}
\end{equation}
The $p^{2}$ terms are consistent with the bosonic nature of the excitations.

Eq.(\ref{mg2}) and Eq.(\ref{mg3}) give us the masses of the pseudoscalar and scalar excitations from the curvature of the energy momentum relation
\begin{equation}
\frac{1}{2m^{P}_{*}e_{L}^{2}a^{3}}=(4\epsilon^2+64\epsilon^4)
\label{mg4}
\end{equation}
\begin{equation}
\frac{1}{2m^{S}_{*}e_{L}^{2}a^{3}}=(-4\epsilon^2+72\epsilon^4)
\label{mg5}
\end{equation}

\section{The chiral condensate}
\setcounter{equation}{0}

In the continuum Schwinger model, the phenomenon 
of dynamical symmetry breaking of the chiral symmetry, 
is due to the anomaly. 
The order parameter is the mass operator 
$M(x)=\overline{\psi}(x)\psi(x)$ 
which acquires a non zero vacuum expectation value, 
giving rise to the chiral condensate
\cite{hoso}
\begin{equation}
\chi_{c}=<\overline{\psi}(x)\psi(x)>
=-\frac{e^{\gamma}}{2\pi}m_{c}=-\frac{e^{\gamma}}{2\pi}\frac{e_{c}}{\sqrt{\pi}}
\label{cond}.
\end{equation}

In this section we compute the lattice chiral condensate 
to the fourth order in perturbation theory.
In the staggered fermion formalism (having put particles on even sites and antiparticles on odd sites), one has 
\begin{equation}
\overline{\psi}(x)\psi(x) \longrightarrow \frac{(-1)^{x}}{a}(\psi^{\dagger}_{x}\psi_{x}-\frac{1}{2})
\end{equation}
The lattice chiral condensate may be obtained by considering
the mass operator
\begin{equation}
M=-\frac{1}{Na}\sum_{x=1}^{N}(-1)^{x}\psi^{\dagger}_{x}\psi_{x}
\label{mmm}
\end{equation}
where the extra minus sign is there to give the same sign to the lattice and continuum chiral condensates (see footnote on page 1), and evaluating the expectation value of (\ref{mmm}) on the perturbed states $|p_{\psi}>$ generated by applying
$H_{h}$ to $|\psi>$. One has
\begin{equation}
|p_{\psi}>=|\psi>+\epsilon|p_{\psi}^{1}>+\epsilon^{2}|p_{\psi}^{2}>
\end{equation}
where
\begin{eqnarray}
|p_{\psi}^{1}>=-\frac{1}{\Delta}H_{h}|\psi>\\
|p_{\psi}^{2}>=\frac{\Pi_{\psi}}{2\Delta^{2}}H_{h}H_{h}|\psi>.
\end{eqnarray}
The lattice chiral condensate is then given by
\begin{equation}
\chi_{L}=\frac{<p_{\psi}|M|p_{\psi}>}{<p_{\psi}|p_{\psi}>}=\frac{<\psi|M|\psi>+\epsilon^{2}<p_{\psi}^{1}|M|p_{\psi}^{1}>+\epsilon^{4
}<p_{\psi}^{2}|M|p_{\psi}^{2}>}{<\psi|\psi>+\epsilon^{2}<p_{\psi}^{1}|p_{\psi}^{1}>+\epsilon^{4}<p_{\psi}^{2}|p_{\psi}^{2}>}
.
\end{equation}
One gets the following expressions for the wave functions
\begin{eqnarray}
<\psi|\psi>&=&1\\
<p_{\psi}^{1}|p_{\psi}^{1}>&=&\frac{N}{\Delta^{2}}\\
<p_{\psi}^{2}|p_{\psi}^{2}>&=&\frac{N(N-3)}{2\Delta^{4}}
\end{eqnarray}
and for the $M$ operator
\begin{eqnarray}
<\psi|M|\psi>&=&-\frac{1}{2a}\\
<p_{\psi}^{1}|M|p_{\psi}^{1}>&=&-\frac{1}{\Delta^{2}a}(\frac{N}{2}-2)\\
<p_{\psi}^{2}|M|p_{\psi}^{2}>&=&-\frac{1}{4\Delta^{4}a}(\frac{N}{2}-4)\cdot 2\cdot (N-3).
\end{eqnarray}
To the fourth order in $\epsilon$, the lattice chiral condensate is given by 
\begin{equation}
\chi_{L}=-\frac{1}{a}(\frac{1}{2}-
\frac{2}{\Delta^{2}}\epsilon^{2}
+\frac{6}{\Delta^{4}}\epsilon^{4})=-\frac{1}{a}(\frac{1}{2}-32\epsilon^{2}+1536\epsilon^{4}).
\label{lchi}
\end{equation}

\section{Lattice versus continuum}
\label{lvc}
\setcounter{equation}{0}

In this section we want to extract 
some physics from the lattice results we obtained; to do this, one should compare the answer 
of the strong coupling analysis of the lattice theory with the exact results of the continuum model.
For this purpose one should extrapolate the strong 
coupling expansion derived under the assumption that the parameter 
$z=\frac{t^{2}}{4e_{L}^{4}a^{4}}\ll1$ to the region in which 
$z\gg1$; this region corresponds to the continuum theory since 
$e_{L}^{4}a^{4}\longrightarrow 0$, $z\longrightarrow \infty$. 
To make this 
extrapolation possible, it is customary 
\cite{pad} to make use of Pad\'e approximants, 
which allow to extrapolate a series expansion beyond 
the convergence radius. 
Strong coupling perturbation theory improved with Pad\'e approximants should be compared with the continuum theory.

As we shall see, the gauge invariant strong coupling expansion we propose,
provides a very accurate estimate of the observables of the continuum theory, 
already at the second order in powers of $z$. This strongly suggests that,
expanding around the gauge, parity and charge conjugation invariant ground state $|\psi>$, leads 
to a perturbative series converging
to the continuum theory faster then the one used in \cite{kog,sus,ham}.  

We first compute 
the ratio between the continuum 
value of the meson mass $m_c =e_c/\sqrt{\pi}$ and the lattice coupling constant 
$e_L$, 
by equating the lattice chiral condensate (\ref{lchi}), to its continuum 
counterpart (\ref{cond}) 
\beq
\frac{1}{a} \left(\frac{1}{2}-32 z+1536z^2\right)=- \frac{e^{\gamma}}{2\pi}
m_c\ \ .
\label{lmecm}
\eeq
Of course, eq.(\ref{lmecm}) is true only when Pad\'e approximants are used, 
since - as it stands - the l.h.s. 
holds only for $z\ll 1$, while the r.h.s. provides the value of the chiral 
condensate to be obtained when $z\simeq \infty$.
Using the relation
\begin{equation}
a=(\frac{t^{2}}{4z})^{\frac{1}{4}}\frac{1}{e_{L}}
\end{equation}
one gets from (\ref{lmecm})
\begin{equation}
\frac{m_{c}}{e_{L}}=(\frac{4z}{t^{2}})^{\frac{1}{4}}\frac{2\pi}{e^{\gamma}}(\frac{1}{2}-32z+1536z^2)
\label{fme1}
\end{equation}
As in ref. \cite{sus}, due to the factor $z^{\frac{1}{4}}$, the fourth power of (\ref{fme1}) should be considered in order to construct a 
non-diagonal Pad\'e approximant. Since the strong coupling expansion has been carried out up to second order in $z$, one is allowed to construct only the 
$[0,1]$ Pad\'e approximant for the polynomial written in (\ref{fme1}). One gets
\begin{equation}
(\frac{m_{c}}{e_{L}})^{4}=\frac{1}{4t^2}(\frac{2\pi}{e^{\gamma}})^{4}\frac{z}{1+256z}
\end{equation}
Taking the continuum limit $z\longrightarrow \infty$ one has
\begin{equation} 
(\frac{m_{c}}{e_{L}})^{4}=\frac{\pi^{4}}{64e^{4\gamma}}\frac{1}{t^{2}}
\label{fme2}
\end{equation}
Next we compute the same ratio by equating the pseudoscalar mass gap given in (\ref{mga}) to its continuum counterpart $m_c$
\beq
e_L^2 a \left(\frac{1}{4}+8z-576 z^2\right)= m_c\ \ .
\label{mcecc}
\eeq
Again eq.(\ref{mcecc}) is true only when Pad\'e approximants are used.

Dividing both sides of eq.(\ref{mcecc}) by $e_{L}$ and taking into account 
that 
\begin{equation}
e_{L}a=(\frac{t^{2}}{4z})^{\frac{1}{4}},
\label{ez}
\end{equation}
one gets 
\begin{equation}
\frac{m_{c}}{e_{L}}=(\frac{t^{2}}{4z})^{\frac{1}{4}}
(\frac{1}{4}+8z-576z^{2})\ ,
\label{moe}
\end{equation}
Taking the fourth power, as we did for the chiral condensate equation, and constructing the $[1,0]$ Pad\'e approximant for the r.h.s. of eq.(\ref{moe}) 
one gets
\begin{equation}
(\frac{m_{c}}{e_{L}})^{4}=\frac{t^{2}}{4z}(\frac{1}{256}+\frac{z}{2})
\end{equation} 
One may now take the limit  $z\to \infty$, obtaining
\begin{equation}
(\frac{m_{c}}{e_{L}})^{4}=\frac{t^{2}}{8}\ .
\label{lm1}
\end{equation}
Equating eq.(\ref{fme2}) and eq.(\ref{lm1}) one gets an equation for $t$ which gives
\begin{equation}
t=\frac{\pi}{8^{\frac{1}{4}}e^{\gamma}}=1.049
\end{equation}
which lies $4.9\%$ above the exact value. It is conforting to see that the lattice theory gives a light velocity very close to 
(but greater than ) $1$. 
Putting this value of $t$ in eq.(\ref{fme2}) or eq.(\ref{lm1}) one has
\begin{equation}
\frac{m_{c}}{e_{L}}=0.609
\end{equation}
which lies $7.9\%$ above the exact value $\frac{1}{\sqrt{\pi}}$.

It is worth to stress that we can reproduce very well the continuum results, even if we use just first order (in z) results of the 
strong coupling perturbation theory.

A second way \cite{kog} to extrapolate the polinomial in (\ref{fme1}) is to equate it to the ratio
\begin{equation}
\frac{(1+bz)^{x}}{(1+az)^{x+\frac{1}{4}}}=1-64z+3072z^{2}
\end{equation}
so that one can take the limit $z\longrightarrow \infty$ in (\ref{fme1}). Setting $t=1$ one gets
\begin{equation}
\frac{m_{c}}{e_{L}}=\frac{\sqrt{2}\pi}{e^{\gamma}}\frac{b^x}{a^{x+\frac{1}{4}}}
\label{1m}
\end{equation}
It is worth to say that this extrapolation works for a large range of values of the parameter $x$ between $0$ and $6$ giving an answer that lies between $6.9\%$ below and $10\%$ above the exact answer. In particular for $x=1$ one gets $\frac{m_{c}}{e_{L}}
=0.596$ which lies $5.6\%$ above the exact value. For $x=\frac{3}{4}$ one gets $\frac{m_{c}}{e_{L}}=0.593$ wich differs from the continuum value of the $5.1\%$. For $x=\frac{1}{10}$ one has only $0.7\%$ of error.
In the same way one can extrapolate the polinomial in (\ref{moe}) 
\beq
\frac{(1+ b z)^{x+\frac{1}{4}}}{(1+az)^x}=1+32 z-2304 z^2
\eeq
so that one can then take the limit $z\to\infty$ in (\ref{moe}).
Setting $t=1$ one gets
\beq
\frac{m_c}{e_L}=\frac{1}{4^{\frac{5}{4}}}\left( b \right)^{x+\frac{1}{4}}
\frac{1}{a^{x}}
\label{617}
\eeq
for $x=1$ one gets $\frac{m_{c}}{e_{L}}=0.617$ which lies $9\%$ above the exact value, against the $30 \%$ error of ref.\cite{kog}. For $x=0.5$ one has  
$\frac{m_{c}}{e_{L}}=0.595$ which lies only $5.4\% $ above the exact answer.

In the continuum Schwinger model the ratio between the masses of the scalar
and the pseudoscalar particles, $m^S/m^P$, equals 2, since $m^S$ is the lowest eigenvalue
of the continuum mass spectrum, which starts at $2 m^P$ \cite{sus}.
We shall now evaluate this ratio using the lattice expression of $m^P$ and $m^S$ 
(\ref{mga},\ref{mgb}). $m^S/m^P$ is given by
\beq
\frac{m^S}{m^P}=\frac{\frac{1}{4}+ 24 z - 1600 z^2}
{\frac{1}{4}+ 8 z - 576 z^2}\ \ .
\label{rm}
\eeq
Expanding the r.h.s. of (\ref{rm}) in power series of $z$ 
\beq
\frac{m^S}{m^P}=1+64 z-6144 z^2\ \ ,
\eeq
one may perform the $[1,1]$ Pad\'e approximant
\beq
\frac{m^S}{m^P}=\frac{1+ 160 z}
{1+96 z}\ \ .
\eeq
For $z\to\infty$
\beq
\frac{m^S}{m^P}=1.67
\eeq
which lies $16\%$ below the continuum value. 
Our result coincides with the one obtained in \cite{sus}
at the same order in the perturbative expansion.

Taking the $[1,1]$ Pad\'e approximants of eq.(\ref{mg4}) and eq.(\ref{mg5}) one can compute in an independent way the ratio between the masses of 
the scalar and pseudoscalar particles
\begin{equation}
\frac{m_{*}^{S}}{m_{*}^{P}}=-\frac{1+18z}{1-16z}
\end{equation}
For $z\longrightarrow \infty$
\begin{equation}
\frac{m_{*}^{S}}{m_{*}^{P}} =1.125
\end{equation}
which lies $43\%$ below the exact value. This means that one should go to the next order in perturbation theory in eq.(\ref{mg4}) and eq.(\ref{mg5}) 
to reproduce better the continuum limit, since in eq.(\ref{mg4}) and eq.(\ref{mg5}) one has no zeroth order term.
Another test (also suggested in \cite{sus}) to check the validity of the lattice 
computations, may be performed by computing the quantity 
\beq
D=\frac{z^{3/4}}{m^P} \frac{d}{dz}\left(z^{1/4} m^P\right)=
\frac{1}{4} + \frac{z}{m^P}\frac{d m^P}{d z}\ .
\eeq
This should equal $1/4$ if the lattice theory has to reproduce 
the continuum result.

Using (\ref{mga}) one gets
\beq
D=z\frac{8-1152 z}{1/4+8z}\simeq 32 z(1-176 z)\ .
\eeq
Constructing the $[0,1]$ Pad\'e approximant and taking the $z\to \infty$ 
limit, one has
\beq
D=z\frac{32}{1+176 z}\stackrel{z\to\infty}{\longrightarrow}=\frac{2}{11}=0.182
\eeq
which lies $27\%$ below the desired $0.25$.
This coincides with the result obtained in \cite{sus}, 
using a strong coupling expansion up to the order $z^4$.
The agreement of our results with the continuum theory is very encouraging, since 
only terms up to the order $z^2$
have been used in the gauge invariant strong coupling expansion proposed 
in this paper.

\section{Concluding remarks}
\setcounter{equation}{0}

Our paper is aimed at constructing an improved strong coupling 
expansion for the lattice Schwinger model. We show that, gauge 
invariance together with the discrete symmetries of the model, force 
the ground state to be a N\'eel state.
Moreover gauge invariance 
requires the ground state to have electric fields related to the 
charge density operator by the Gauss's law. 
Thus, for large $e_{L}^{2}$, the ground state energy 
is of order $e_{L}^{2}$ rather 
than $\frac{1}{e_{L}^{2}}$ \cite{kog,sus}; as a consequence,
the strong coupling expansion we construct
provides a very accurate extrapolation to the 
continuum theory already at the second order in 
$z=\frac{t^{2}}{4e_{L}^{4}a^{4}}$ . Furthermore, 
the Coulomb energy of elementary excitations is also affected, 
since $-$ besides the Coulomb self 
energy $-$ there is also an interaction energy with the charge of the ground state. 

Strong coupling $QED$ with $N_{L}$ flavors of staggered fermions has been 
shown to be equivalent $-$
in any space dimension D $-$ to a quantum $SU(N_{L})$ antiferromagnet 
\cite{dssl}. For the strongly coupled  lattice Schwinger model, one has  
equivalence with a one-dimensional antiferromagnetic Ising spin chain
with a Coulomb long range interaction, which admits the N\'eel state as 
an exact ground state.
Due to this, in our approach, chiral symmetry 
is broken spontaneously in the lattice model already at 
zero-th order in $z$. We expect that chiral symmetry breaking should persist for all
coupling, since the critical coupling for $D=1$ should be at $e_L^2=0$.
 We conjecture that, for small $e^{2}$, this behavior 
is a manifestation of a Peierls instability
$-$ the tendency of a one dimensional Fermi gas to form a gap at the Fermi 
surface. This happens with any infinitesimal interaction. 

In the
continuum theory the Schwinger model exhibits 
the Nambu-Goldstone phenomenon: 
global chiral symmetry is spontaneously broken, but no Goldstone boson
appears in the spectrum since the local current is 
anomalous \cite{cole}. On the lattice there is no anomaly
due to the Nielsen-Ninomiya theorem \cite{niels}. 
This is obviously true in the staggered fermion formalism since
\begin{equation}
Q_{5}=\sum_{x}(\psi_{x}^{\dag}\psi_{x+1}+\psi_{x+1}^{\dag}\psi_{x})
\end{equation}  
and
\begin{equation}
\dot{Q_{5}}=i[H_{S},Q_{5}]=0\ \ ,
\end{equation}
which leads to no net-production of particles with a given handedness.

Our analysis shows that in the strong coupling limit, chiral 
symmetry is spontaneously broken at all orders,
since the states $|\psi>$ and $|\chi>$ mix
only at a perturbative order comparable with the volume of the system.
Being chiral
symmetry replaced by the discrete  
symmetry representing invariance under 
tranlations by one lattice site, effects related to the
breaking of the chiral symmetry on the lattice 
should come from the coupling between the two sublattices. 
This is manifest in our strong coupling calculations.

It will be interesting to extend our gauge invariant strong coupling analysis 
to include at least the case where $N_{L}=2$; in this case, it is expected
 \cite{dssl} that the effective Hamiltonian for the model is the antiferromagnetic Heisenberg Hamiltonian. 
This analysis could provide a 
framework to analyse the properties of the spinon and holon excitations of quantum condensed 
 matter models in the context of gauge theories.

\section{Appendix A}
\setcounter{equation}{0}

The doubling problem of lattice fermions and a solution to cure it, are reviewed.
Let us consider the massless Dirac equation in $1+1$ dimensions in the continuum
\begin{equation}
i\dot{\psi}=-i\alpha \partial_{x}\psi=-i\gamma_{5}\partial_{x}\psi
\label{dire}
\end{equation}
with  $\psi=\left(\begin{array}{c}
\psi_{1}\\
\psi_{2}
\end{array}\right)$
and $\alpha=\gamma_{5}=\sigma_{1}$.  
Choosing for the plane waves 
\begin{equation}
\psi_{\pm}(x,t)=e^{-i(kx-Et)} \chi_{\pm}
\end{equation}
with
\begin{equation}
\gamma_{5}\chi_{\pm}=\pm \chi_{\pm}
\end{equation}
the dispersion relation is 
\begin{equation}
E(k)=\pm k\quad,\quad -\infty<k<\infty
\end{equation}
and the excitations are left- and right-handed particles and antiparticles.

Let us consider the naively latticized Dirac equation. 
We place the spinor $\psi=\left(\begin{array}{c}
\psi_{1}\\
\psi_{2}
\end{array}\right)$ on each site of a spatial lattice and change the partial derivative $\partial_{x}$ with a finite difference
\begin{equation}
i\dot{\psi}(x)=-\frac{i}{2a}\gamma_{5}[\psi(x+1)-\psi(x-1)]
\end{equation}
The dispersion relation is now 
\begin{equation}
E(k)=\pm \frac{\sin (ka)}{a}\ .
\label{disp}
\end{equation}
This is a relativistic 
dispersion relation for $ka\ll 1$
\begin{equation}
E(k)\simeq \pm k+O(k^{3}a^{2})
\end{equation}
and for $ka=\pi -k'a$, with $k'a\ll 1$,
\begin{equation}
E(k)\simeq \mp k'+O(k'^{3}a^{2})
\end{equation}
Let us now turn to the doubling problem of lattice fermions. 
There are two two-component Dirac particles in the continuum limit, 
and the total chiral charge of the fermions is zero. Consider, 
for example, the excitations associated with the field $\psi_{+}$
on the lattice. 
There are right-movers ($k\simeq 0$) and left-movers ($k\simeq \pi$), and since chirality (helicity for particles) is
 just velocity in $1+1$ dimensions this lattice field describes a pair of fermions with net chirality zero. The doubling problem of lattice fermions 
 is related to the continuous chiral symmetry \cite{niels}. 
In the literature one can find three methods to cure this problem: Wilson fermions, SLAC derivatives and staggered fermions.
The Wilson fermion formalism lifts the energy at the edge of the Brillouin zone. The SLAC formalism uses a nonlocal lattice derivative.
In what follows we shall review only the staggered fermion formalism
that we have adopted in this paper.
In this formalism one eliminates the unwanted fermion modes by reducing the Brillouin zone, $i.e.$ by doubling the effective lattice spacing. 
In the case of a $d$-dimensional lattice, one subdivides it into elementary $d$-dimensional hypercubes of unit length. At each site within a given hypercube 
one places a different degree of freedom and repeat this structure periodically throughout the lattice. Since there are $2^{d}$ sites within a hypercube, 
but only $2^{\frac{d}{2}}$ components of a Dirac field, we need $2^{\frac{d}{2}}$ different Dirac fields to reduce the Brillouin zone by a factor of $\frac{1}{2}$. 
The lattice Schwinger model in the Hamiltonian formalism \cite{hwg} with one species of staggered fermions \cite{ssf} \cite{kkg} represents the very particular 
example in which the number of sites of the ``hypercube" is equal to the number of components of the Dirac field.

The main idea is to reduce the number of degrees of freedom by using a single component Fermi field $\psi_{x} $ on each site of the lattice
\begin{equation}
\{\psi_{x},\psi_{y}\}=0\quad,\quad \{\psi^{\dag}_{x},\psi_{y}\}=\delta_{xy}
\end{equation}
so that the lattice Dirac Hamiltonian reads
\begin{equation}
H_{D}=-\frac{it}{2a}\sum_{x}(\psi_{x+1}^{\dag}\psi_{x}-\psi_{x}^{\dag}\psi_{x+1})
\label{dirl}
\end{equation}
and the equations of motion are
\begin{equation}
\dot{\psi}_{x}=-i[H_{D},\psi_{x}]=\frac{t}{2a}[\psi_{x+1}-\psi_{x-1}]
\end{equation}
If one decomposes the lattice into 
even and odd sublattices 
(characterized by $x$ even and odd), one can identify a single two component 
Dirac spinor by associating the upper component 
with even sites and the lower component with odd sites (or vice versa). 
In this way the equations of 
motion become
\begin{equation}
\dot{\psi_{1}}(x)=\frac{1}{2a}[\psi_{2}(x+1)-\psi_{2}(x-1)]
\label{direla}
\end{equation}
\begin{equation}
\dot{\psi_{2}}(x)=\frac{1}{2a}[\psi_{1}(x+1)-\psi_{1}(x-1)]\ .
\label{direlb}
\end{equation}
When one compares eq.(\ref{direla})and eq.(\ref{direlb}) with the
the continuum Dirac equation,
it is easy to understand how the staggered fermion 
formalism works:
 it avoids the species doubling problem by doubling the lattice spacing and so 
 halving the size of the Brillouin zone.

Let us now discuss the symmetries of the 
lattice Dirac Hamiltonian (\ref{dirl}). 
It is invariant under translations of the spatial lattice 
by an even number of sites. 
Translation by two lattice spacings is the ordinary continuum translation, 
whose generator is 
\begin{equation}
p=-i\int dx \psi^{\dag}\partial _{x} \psi =-i\int dx(\psi_{1}^{\dag}\partial_{x}\psi_{1}+\psi_{2}^{\dag}\partial_{x}\psi_{2}) 
\end{equation}
and it does not mix upper and lower spinor components. 
Consequently, one expects the lattice generator not to mix the two sublattices
\begin{equation}
p=-i\sum_{x}(\psi_{x+2}^{\dag}\psi_{x}+\psi_{x}^{\dag}\psi_{x+2})
\end{equation}

The Hamiltonian (\ref{dirl}) is also invariant under translations by an odd number of sites. The lattice generator of this symmetry is
\begin{equation}
T_{5}=\sum_{x}(\psi_{x}^{\dag}\psi_{x+1}+\psi_{x+1}^{\dag}\psi_{x})
\end{equation}
which in the continuum would be $ T_{5}=\int dx \psi^{\dag}\gamma_{5}\psi  $. 
The matrix $\gamma_5$ applied to the bispinor interchanges upper and lower component, this on the lattice corresponds to
translation by one lattice site.
Therefore, the staggered fermion formalism has a discrete 
chiral symmetry, corresponding to chiral a rotation of
$\frac{\pi}{2}$ radiants.

Let us consider now the naive continuum limit of the 
lattice Dirac Hamiltonian (\ref{dirl}). 
We consider an elementary ``plaquette" of the lattice 
as the segment of length $a$ with two sites generated by taking a site of even coordinates and adding to it the lattice spacing $a$. 
We also decompose the lattice into two sublattices generated by taking a site of the elementary segment and translating it by all even 
multiples of the lattice spacing. We label the fermions which reside on the sublattice of each of the site of the elementary segment as 
$\psi_{\alpha}(x)$ with $\alpha=0$ if one is on the first site and $\alpha=1$ if one is on the second site. In momentum space the Hamiltonian is
\begin{equation}
H_{D}=\int_{\Omega_{B}} dk \psi_{\alpha}^{\dag}(k)\Gamma^{1}_{\alpha \beta}\sin ka \psi_{\beta}(k)
\end{equation}
where $\Omega_{B}=\{k: -\frac{\pi}{2a}<k\leq \frac{\pi}{2a}\}$ is the Brillouin zone of the even sublattice. 
The momentum space fermions have the anticommutator 
\begin{equation}
\{\psi_{\alpha}(k),\psi_{\beta}^{\dag}(k')\}=\delta_{\alpha \beta}\delta_{k,k'}
\end{equation}
and the Dirac tensor is
\begin{equation}
\Gamma^{1}_{\alpha,\beta}=\sigma^{1}_{\alpha \beta} 
\end{equation}
The spectrum of the Dirac operator is
\begin{equation}
E(k)=\pm\frac{\sin ka}{a}
\end{equation}
and one notes that, to set the Fermi level of the fermions at the degeneracy point where the two branches of the spectrum meet, 
it is necessary that the fermion states are exactly half filled. This is also required for charge-conjugation invariance, or particle-hole symmetry of
the vacuum state.

\section{Appendix B}
\setcounter{equation}{0}

In this appendix the energies of the excitations generated 
by applying the hopping Hamiltonian $H_h$ to the ground state $|\psi>$, 
are evaluated.
The action of $H_h$ on the state $|\psi>$ generates a dimer-antidimer pair.
All the states with one dimer-antidimer pair,
and with two dimer-antidimer pairs 
(obtained applying $H_{h}$ twice), have the same energy 
gap with respect to the ground state.
However, not all the states obtained applying $(H_h)^n$, with $n\ge3$,
to $|\psi>$ have the same energy. 
There are, for example, two possible energies, corresponding to a gap of 
$3\Delta=3e^2_L a/4$ and $5\Delta$, for the 
states with three dimer-antidimer pairs. 

The states with a dimer-antidimer pair are given by
\begin{equation}
|S^{1}_{R,L}>=H_{R,L}(y)|\psi>.
\end{equation}
Applying on these states the unperturbed Hamiltonian we get
\begin{eqnarray}
H_{u}|S^{1}_{R,L}>&=&\frac{e^{2}_{L}a}{2}\sum_{x}
(E_{x}\pm\delta_{xy})^{2}|\psi>=\nonumber\\
&=&[E_{\psi}^{0}+\frac{e^{2}_{L}a}{2}\sum_{x}(-\frac{(-1)^{x}}{2}\delta_{xy}+\delta_{xy})]|\psi>
\label{bba}
\end{eqnarray}
where $y$ must be even if we apply $H_{R}$ and must be odd 
if we apply $H_{L}$. We thus see that all the states $|S_{R,L}^{1}>$ have
a mass gap of $\Delta$.

The states with two dimer-antidimer pairs are given by
\begin{equation}
|S^{2}_{(R,L)(R,L)}>=H_{R,L}(z)H_{R,L}(y)|\psi>
\end{equation}
\begin{equation}
H_{u}|S^{2}_{(R,L)(R,L)}>=\frac{e^{2}_{L}a}{2}\sum_{x}(E_{x}\pm \delta{xz}\pm 
\label{bbb}
\delta_{xy})^{2}|\psi>
\end{equation}
It is easy to show that all these states have an energy gap of $2\Delta$.

Based on (\ref{bba}) and (\ref{bbb}) one might expect that the states with three dimer-antidimer pairs have 
an energy gap of $3\Delta$.  This is actually true for all the states of this type
except for the 2 states with 
three particles followed or preceeded by three antiparticles. These states have an 
energy gap of $5\Delta$. If one considers the state
\begin{equation}
|S^{3}_{RRR}>=H_{R}(w)H_{R}(z)H_{R}(y)|\psi>,
\end{equation}
one easily obtains that
\begin{eqnarray}
H_{u}|S_{RRR}^{3}>&=&\frac{e^{2}_{L}a}{2}\sum_{x}(E_{x}+\delta_{xw}+\delta_{xz}+\delta_{xy})^{2}|  \psi>=\nonumber\\
&=&[E_{\psi}^{0}+\frac{e^{2}_{L}a}{2}\sum_{x}(-\frac{(-1)^{x}}{2}\delta_{xw}-\frac{(-1)^{x}}{2}\delta_{xz}-\frac{(-1)
^{x}}{2}\delta_{xy}+\nonumber\\
&+&2\delta_{xw}\delta_{xz}+2\delta_{xw}\delta_{xy}+2\delta_{xz}\delta_{xy}+
\nonumber\\
&+&\delta_{xw}+\delta_{xz}+\delta_{xy})]
|\psi>\ .
\end{eqnarray}
To generate a state with three antiparticles followed by three particles,
one must have
\begin{eqnarray}
z=y-2\\
w=y-1
\end{eqnarray}
The energy gap for this state is $5\Delta$. 
The same result is obtained for the
state $|S^{3}_{LLL}>=H_{L}(w)H_{L}(z)H_{L}(y)|\psi>$
with $z=y+3$ and $w=y+2$, which contains three particles followed by three 
antiparticles.

The reason for  these different behaviours,
lies on the fact that the commutation relation between 
$H_{0}$ and $H_{h}$ is
\begin{eqnarray}
[H_{0},H_{h}]=\frac{e^{2}_{L}a}{4}H_{h}+
\frac{e^{2}_{L}t}{8}(\sum_{j\geq k}-\sum_{k\geq j})
(\psi^{\dag}_{k+1}e^{iA}\psi_{k}-\psi_{k}^{\dag}e^{-iA}\psi_{k+1})\rho_{j}
\label{aaa}
\end{eqnarray}
If the second term in the r.h.s. was absent, all the states 
with $n$ dimer-antidimer pairs would have 
an energy gap of $n\Delta$.

\section{Appendix C}
\setcounter{equation}{0}

In this appendix we provide the matrix elements 
needed in the strong coupling expansion expression of the mass gap.
The six matrix elements that arise in 
the computation of the second order bosonic excitation energies, are 
\begin{eqnarray}
M_{1}&=&<\psi|H_{R}(y)H_{R}\Lambda_{\theta}H_{L}H_{L}(x)|\psi>=\nonumber\\
&=&-\frac{1}{\Delta}(\frac{N}{2}-1)\delta_{xy}-\frac{1}{\Delta}(1-\delta_{xy})\\
M_{2}&=&<\psi|H_{L}(y)H_{L}\Lambda_{\theta}H_{R}H_{R}(x)|\psi>=\nonumber\\
&=&-\frac{1}{\Delta}(\frac{N}{2}-1)\delta_{xy}-\frac{1}{\Delta}(1-\delta_{xy})\\
M_{3}&=&<\psi|H_{R}(y)H_{L}\Lambda_{\theta}H_{R}H_{L}(x)|\psi>=\nonumber\\
&=&-\frac{1}{\Delta}(\frac{N}{2}-3)\delta_{xy}+\frac{1}{\Delta}(1-\delta_{xy})\\
M_{4}&=&<\psi|H_{L}(y)H_{R}\Lambda_{\theta}H_{L}H_{R}(x)|\psi>=\nonumber\\
&=&-\frac{1}{\Delta}(\frac{N}{2}-3)\delta_{xy}+\frac{1}{\Delta}(1-\delta_{xy})\\
M_{5}&=&-<\psi|H_{R}(y)H_{L}\Lambda_{\theta}H_{L}H_{R}(x)|\psi>=\nonumber\\
&=&-\frac{1}{\Delta}(\delta_{y,x-1}+\delta_{y,x+1})\\
M_{6}&=&-<\psi|H_{L}(y)H_{R}\Lambda_{\theta}H_{R}H_{L}(x)|\psi>=\nonumber\\
&=&-\frac{1}{\Delta}(\delta_{y,x-1}+\delta_{y,x+1})
\label{mi}
\end{eqnarray}
where $M_{1}$ and $M_{3}$ are defined for $x$ and $y$ odd, $M_{2}$ and $M_{4}$ for $x$ and $y$ even, $M_{5}$ for $x$ even and $y$
odd, and $M_{6}$ for $x$ odd and $y$ even.

To compute the fourth order energy gap one needs also 
the matrix elements $M_{i}'$, $i=1,...,6$ obtained by replacing $\Lambda_{\theta}$ with $\Lambda_{\theta}^{2}$ in $M_{i}$
\begin{eqnarray}
M_{1}'=M_{2}'=M_{3}'=M_{4}'&=&\frac{1}{\Delta^{2}}(\frac{N}{2}-1)
\delta_{xy}+\frac{1}{\Delta^{2}}(1-\delta_{xy})\\
M_{5}'=M_{6}'&=&-\frac{1}{\Delta^{2}}(\delta_{y,x-1}
+\delta_{y,x+1})-\frac{2}{\Delta^{2}}(1-\delta_{y,x-1}-\delta_{y,x+1})
\end{eqnarray}
with the same constraints on $x$ and $y$ as before.

One must then  evaluate the twenty matrix elements arising when three $H_R$ 
and three $H_L$ are combined in all possible ways in the expression
 in the first line of eq.(\ref{et4})
\begin{equation}
<\theta(y)|(H_{L}-H_{R})\Lambda_{\theta}(H_{L}-H_{R})\Lambda_{\theta}
(H_{R}-H_{L})\Lambda_{\theta}(H_{R}-H_{L})|\theta(x)>
\end{equation}
Since hermiticity of the Hamiltonian requires that the elements 
obtained changing $H_{R}$ with $H_{L}$ are equal 
one has only to compute the ten matrix elements, 
\begin{eqnarray}
M_{1}''&=&\frac{1}{N}\sum_{x,y}M''_{1}(x,y)=\frac{1}{N}\sum_{x,y}<\psi|H_{R}(y)H_{R}\Lambda_{\theta}H_{R}\Lambda_{\theta}H_{L}\Lambda_{\theta}H_{L}H_{L}(x)|\psi>\\
M_{2}''&=&\frac{1}{N}\sum_{x,y}M''_{2}(x,y)=\frac{1}{N}\sum_{x,y}<\psi|H_{R}(y)H_{L}\Lambda_{\theta}H_{L}\Lambda_{\theta}H_{R}\Lambda_{\theta}H_{R}H_{L}(x)|\psi>\\
M_{3}''&=&\frac{1}{N}\sum_{x,y}M''_{3}(x,y)=\frac{1}{N}\sum_{x,y}<\psi|H_{R}(y)H_{R}\Lambda_{\theta}H_{L}\Lambda_{\theta}H_{R}\Lambda_{\theta}H_{L}H_{L}(x)|\psi>\\
M_{4}''&=&\frac{1}{N}\sum_{x,y}M''_{4}(x,y)=\frac{1}{N}\sum_{x,y}<\psi|H_{R}(y)H_{R}\Lambda_{\theta}H_{L}\Lambda_{\theta}H_{L}\Lambda_{\theta}H_{R}H_{L}(x)|\psi>\\
M_{5}''&=&\frac{1}{N}\sum_{x,y}M''_{5}(x,y)=\frac{1}{N}\sum_{x,y}<\psi|H_{R}(y)H_{L}\Lambda_{\theta}H_{R}\Lambda_{\theta}H_{R}\Lambda_{\theta}H_{L}H_{L}(x)|\psi>\\
M_{6}''&=&\frac{1}{N}\sum_{x,y}M''_{6}(x,y)=\frac{1}{N}\sum_{x,y}<\psi|H_{R}(y)H_{L}\Lambda_{\theta}H_{R}\Lambda_{\theta}H_{L}\Lambda_{\theta}H_{R}H_{L}(x)|\psi>\\
M_{7}''&=&\frac{1}{N}\sum_{x,y}M''_{7}(x,y)=-\frac{1}{N}\sum_{x,y}<\psi|H_{R}(y)H_{R}\Lambda_{\theta}H_{L}\Lambda_{\theta}H_{L}\Lambda_{\theta}H_{L}H_{R}(x)|\psi>\\
M_{8}''&=&\frac{1}{N}\sum_{x,y}M''_{8}(x,y)=-\frac{1}{N}\sum_{x,y}<\psi|H_{R}(y)H_{L}\Lambda_{\theta}H_{R}\Lambda_{\theta}H_{L}\Lambda_{\theta}H_{L}H_{R}(x)|\psi>\\
M_{9}''&=&\frac{1}{N}\sum_{x,y}M''_{9}(x,y)=-\frac{1}{N}\sum_{x,y}<\psi|H_{R}(y)H_{L}\Lambda_{\theta}H_{L}\Lambda_{\theta}H_{R}\Lambda_{\theta}H_{L}H_{R}(x)|\psi>\\
M_{10}''&=&\frac{1}{N}\sum_{x,y}M''_{10}(x,y)=-\frac{1}{N}\sum_{x,y}<\psi|H_{R}(y)H_{L}\Lambda_{\theta}H_{L}\Lambda_{\theta}H_{L}\Lambda_{\theta}H_{R}H_{R}(x)|\psi>\ .
\end{eqnarray}
where $M_{1}''(x,y),...,M_{6}''(x,y)$ are defined for $x$ and $y$ odd, $M_{7}''(x,y),...,M_{10}''(x,y)$ are defined for $x$ even and $y$ odd.
The results are
\begin{eqnarray}
M_{1}''(x,y)&=&-\frac{1}{2\Delta^{3}}\delta_{x,y}-\frac{1}{4\Delta^{3}}(N-2)(N-4)\delta_{x,y}+\nonumber \\
&-&\frac{1}{4\Delta^{3}}(\delta_{y,x+2}+\delta_{y,x-2})-\frac{1}{\Delta^{3}}(N-4)(1-\delta_{x,y}) \\
M_{2}''(x,y)&=&-\frac{1}{4\Delta^{3}}(N-4)(N-6)\delta_{x,y}\\
M_{3}''(x,y)&=&M_{4}''(x,y)=M_{5}''(x,y)=M_{6}''(x,y)=-\frac{1}{8\Delta^{3}}(N-4)(N-6)\delta_{x,y}+\nonumber \\
&-&\frac{1}{4\Delta^{3}}(N-6)(\delta_{y,x+2}+\delta_{y,y-2})-\frac{1}{4\Delta^{3}}(1-\delta_{x,y}-\delta_{y,x+2}-\delta_{y,x-2})\\
M_{7}''(x,y)&=&M_{8}''(x,y)=M_{9}''(x,y)=M_{10}''(x,y)=\nonumber\\
&=&\frac{1}{2\Delta^{3}}(N-6)(1-\delta_{y,x-1}-\delta_{y,x+1})
\end{eqnarray}
and
\begin{eqnarray}
M_{1}''&=&-\frac{1}{2\Delta^{3}}-\frac{3}{8\Delta^{3}}(N-2)(N-4)\\
M_{2}''&=&M_{3}''=M_{4}''=M_{5}''=M_{6}''=-\frac{1}{8\Delta^{3}}(N-4)(N-6)\\
M_{7}''&=&M_{8}''=M_{9}''=M_{10}''=+\frac{1}{8\Delta^{3}}(N-4)(N-6)
\end{eqnarray}


\begin{thebibliography}{99}
\bibitem{adler}S.L. Adler and W.A. Bardeen , Phys. Rev. 182 (1969) 1517,
               J.S. Bell and R. Jackiw, Nuovo Cimento 51 (1969) 47.   
\bibitem{niels}H.B. Nielsen and M. Ninomiya, Nucl. Phys. B185 (1981) 20, ibid B193 (1981) 173, Phys. Lett. B105 (1981) 219, Phys. Lett. B130 (1983) 389.
\bibitem{hwg}J.B. Kogut and L. Susskind, Phys. Rev. D11 (1975) 395.
\bibitem{ssf}L. Susskind, Phys. Rev. D16 (1977) 3031.
\bibitem{kkg}J.B. Kogut, Rev. Mod. Phys. 55 (1983) 775.
\bibitem{hoso} J.E. Hetrick and Y. Hosotani, Phys. Rev. D38 (1988) 2621, 
J.E. Hetrick, Y. Hosotani and S. Iso hep-th/9502113,
Y. Hosotani, R. Rodriguez,J.E. Hetrick and S. Iso hep-th/9606129,
Y. Hosotani hep-th/9606167 .
\bibitem{scw} J.Schwinger, Phys. Rev. 128 (1962), 2425.
\bibitem{low} J. Lowenstein and J.A. Swieca, Ann. Phys. 68 (1971), 172.
\bibitem{cole} S. Coleman, R. Jackiw and L. Susskind, Ann. Phys. 93 (1975) 267. 
\bibitem{kog}T. Banks, L. Susskind and J. Kogut, Phys. Rev. D13 (1976) 1043.
\bibitem{sus}A. Carroll, J. Kogut, D. K. Sinclair and L. Susskind, Phys. Rev. D13 (1976) 2270.
\bibitem{dssl}M. C. Diamantini, E. Langman, G. W. Semenoff and P. Sodano 
Nucl. Phys. B406 (1993)595.
\bibitem{man} N.S. Manton, Ann. Phys. 159 (1985), 220.
\bibitem{lsm}E. Lieb, T. Schultz and D. Mattis, Ann. Phys. (N.Y.) 16, 407 (1961).
\bibitem{nambu}Y. Nambu, Prog. Theor. Phys. 116 (1474) (1950).
\bibitem{ham} C.J. Hamer, Zheng Weihong, and J. Oitmaa, Preprint hep-lat/9701015.
\bibitem{cteo}S. Coleman Commun. Math. Phys. 31 (1973) 259.
\bibitem{pad}G. A. Baker, Jr., Essential of Pad\'e Approximants (Academic, New York, 1975).
\end{thebibliography}
\end{document}